# Implementation of an Elastic Reconfigurable Optical Add/Drop Multiplexer based on Subcarriers for Application in Optical Multichannel Networks

Faranak Khosravi
*Dep. Electrical Engineering
University of Texas at San Antonio*
San Antonio, USA
Faranak.khosravi@my.utsa.edu

Mehdi Tarhani
*Dep. Electrical Engineering
University of Texas at San Antonio*
San Antonio, USA
mehdi.tarhani@utsa.edu

Shivani Kurle
*Dep. Electrical Engineering
University of Texas at San Antonio*
San Antonio, USA
Shivani.Kurle@my.utsa.edu

Mehdi Shadaram
*Dep. Electrical Engineering
University of Texas at San Antonio*
San Antonio, USA
Mehdi.Shadaram@utsa.edu

**Abstract-** We designed a Reconfigurable Optical Add/Drop Multiplexer (ROADM) based on a subcarrier add/drop node in an optical communication system that is suitable for all kinds of optical multiplexing signals. To achieve this goal, at first, we designed an optical comb generator based on a dual-drive Mach Zehnder. The new ROADM setup is validated by a 100 Gb/s 4-subcarrier. In the final step, we checked the performance of the system in terms of the bit error rate (BER) versus optical signal-to-noise ratio (OSNR) to verify the add/drop operation had been successfully performed at $10^{-9}$ and is suitable to apply in an all-optical multiplexing technique.

***Keywords- ROADM, Optical carrier generator-Multiplexing, Subcarriers, Multichannel networks***

## I. INTRODUCTION

Increased data transfers that lead to optical networks have faced exponential growth during recent decades [1, 2]. Optical networks provide high-capacity telecommunications by using optical components that have the capacity to transfer high volumes of information [3, 4]. As optical network traffic grows, better infrastructure is needed to accommodate the high user volume. Furthermore, cheaper and flexible techniques and methods become more necessary [5, 6]. Modulation and multiplexing techniques are used to design a ROADM that meets these requirements as well as maximizes spectral efficiency [5, 7]. A ROADM is a subsystem that allows adding and/or dropping of wavelengths on a network node [5, 8]. Key properties of the ROADM include automatic power equalization, and well as colorless and directionless capabilities [1, 10]. One of the advantages of colorless directionless contentionless (CDC) ROADMs is the capability to handle their functionality automatically without needing to adjust the power levels manually. The colorless capability has advantages in a ROADM because it simplifies operations, enhances the value of using tunable transponders, and is highly desirable for a network with dynamic traffic [4]. The directionless property is important, especially in the event of failure, as it leads to an add or drop transponder to access the links that enter or exit a ROADM [6, 8]. Also, ROADMs decrease maintenance time and optical equipment to modify optical networks. To the best of our knowledge, we proposed a compact and cost-effective CDC ROADM with a dual-drive Mach-Zehnder

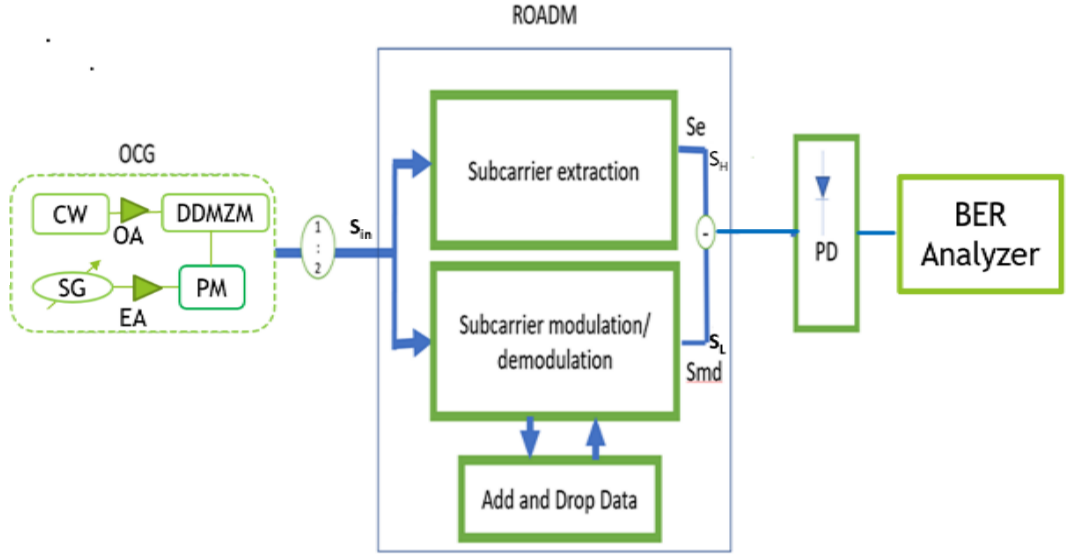

Fig1: Set up of basic system of ROADM

modulator (DDMZM) with optimum splitting ratio. In the previous designs [9] used two single drive Mach Zehnder modulators, while in this research we used only one DDMZM thus making it a more compact and efficient system. In this paper, a ROADM was developed and used to transmit 4- subcarrier signals; the developed technique worked successfully as an add/drop on intermediate nodes in a point-to-point transmission path that increased the flexibility of the system.

## II. METHODOLOGY

Fig. 1 shows different parts of the system structure that used a basic ROADM. The optical comb generator (OCG) is a key part of our cost-efficient ROADM design as it only uses one seed laser instead of many. The OCG comprises a combination of phase modulators (PM) and DDMZM. The Phase modulator in this setup with π/2 phase works by sine wave generator (SG) and PM assist DDMZM in OCGs to generate the comb spectrum of spectral lines [6]. This design generates signals with multiple carriers that make better use of the available optical bandwidth. Once the super-channel has been generated, it undergoes a back-to-back connection to the ROADM, being sampled at its input by a power splitter. A different number of add and drop ports exist in each ROADM node. The spectrum of the input signal of the ROADM is partially converted by the processing block in the lower interferometer pathway, which contains the drop and add subcarrier coherently into the digital domain. Generating the difference between the noisy drop subcarrier and an add subcarrier digitally and converting back to optics yields the complete subcarrier add/drop functionality. The input signal is [11]:

$$s_{in} = \sum_m s_m(t)\, exp\,(j2\pi f_m t + j\varphi_m(t)) \quad (1)$$

where $s_m$ is the mth subcarrier of the input signal, $f_m$ is the frequency of laser, and $\varphi_m$ is the phase noise of laser. Fig. 1 shows the resultant drop signal:

$$s_H(t) = s_e(t)\exp\,(j2\pi(f_i - f_l)t +$$
$$j[\varphi_i(t) - \varphi_l(t)]) \quad (2)$$

which is subtracted from the new subcarrier's modulated waveform which should be replaced, ($f_i$ and $f_l$, are frequencies of the output of OCG). The new subcarrier might be optionally shifted by frequency to finally occur exactly at the optical frequency:

$$s_L(t) = s_{md}(t)\exp\left(j2\pi(f_i - f_l t)\right) \quad (3)$$

The difference signal, given by:

$$s_{HL}(t) = s_H(t) - s_L(t) \quad (4)$$

is modulated onto a part of the same wavelength. A coupler is superimposed onto the optical signal of the fully optical interferometer pathway. Moreover, it is essential to correctly set the interferometer phase difference to obtain the considered interference at the output coupler. Appropriate fiber delays and phase shifters can yield all these circumstances along with tunable digital delays as well as implementing digital phase shifters, which should be controlled by feedback signals [11, 12]:

$$s_{out}(t) = \sum_m s_m(t)e^{(j2\pi f_m t + j\varphi_m(t))} + s_{HL}(t)e^{(j2\pi f_l t + j\varphi L(t))}$$
$$= \sum_{m \neq e} s_m(t)e^{(j2\pi fmt + j\varphi_m(t))} + s_e(t)e^{(j2\pi f_l t + j\varphi L(t))} \quad (5)$$

In this part, the signal providing the subcarrier should be completely substituted by the frequency. The optical field–fills the time/frequency modes of the target subcarrier by entering the add/drop node, which comprises an arbitrary 'signal', including both nonlinear and linear distortions and optical noise. Therefore, during the interferometric drop procedure, optical noise and signal distortions within the signal's frequency modes are suppressed effectively. In addition to the probable bandwidth-narrowing impacts for definite kinds of super-channels, add/drop node concentration concerns are inherently reduced as well. In the final stage, the signal with different wavelengths passes through a photodiode receiver (PD) with a 0.7 A/W responsivity, the output of which is monitored by the BER analyzer.

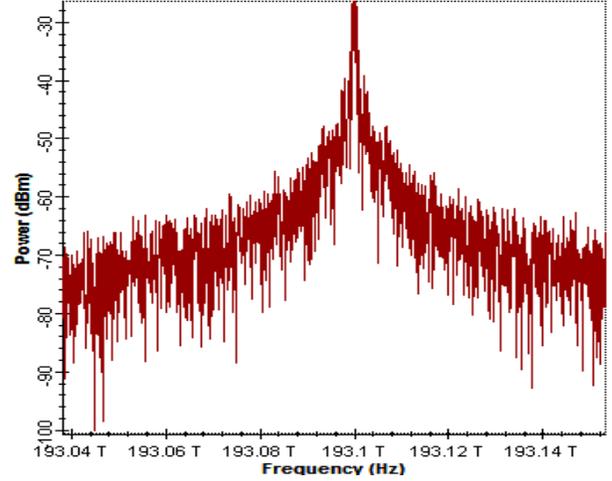

(a)

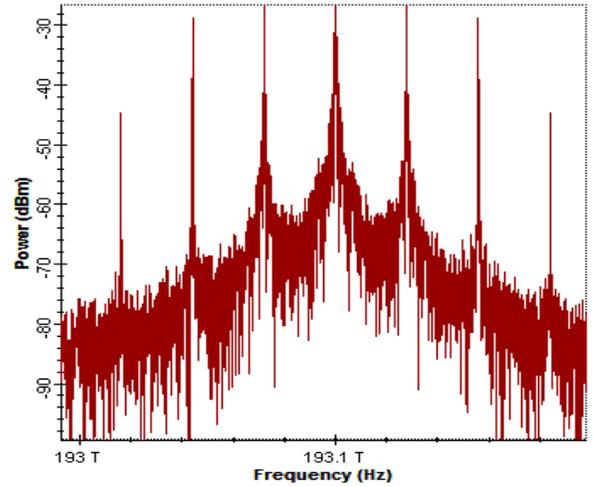

(b)

Fig. 2: (a) Optical spectrum of the source, (b) Optical spectrum of the output of OCG.

### III. RESULTS

Fig. 2 shows the optical spectrum of the source and output of OCG, where the frequency of the CW laser was 193.1 THz. To compensate for losses of the loop, an electrical amplifier with 10 dB gain was used. By comparing Fig. 2(a) and (b), it can be concluded that this technique can produce a set of optical signals that reduces the need for seed lasers.

In the final step, we analyzed the performance of our system in terms of BER versus OSNR to verify if the add/drop operation has been successfully performed.

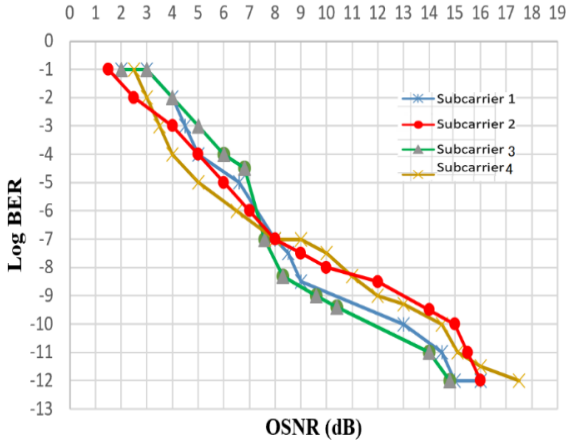

Fig. 3: BER vs. OSNR response in the back-to-back setup for the 4 subcarriers.

Fig. 3 shows that four subcarriers with the same BER under identical OSNR could be obtained in this back-to-back setup. It can thus be concluded that all four subcarriers have the same performance in this system but with added advantages in terms of flexibility and reducing the optical power budget.

## IV. CONCLUSION

This research shows an efficient design of an optical comb generator and a ROADM for all multiplexing techniques. The performance of four subcarriers in the system was compared by BER. The results show that applying this system in optical multichannel networks, such as an elastic optical network, can significantly speed up and reduce the cost for future work.

## REFRENCES